# Hi-throughput gene expression analysis at the level of single proteins using a microfluidic turbidostat and automated cell tracking

G. Ullman[1,2,¤], M. Wallden[1,¤], E. G. Marklund[1], A. Mahmutovic[1], Ivan Razinkov[3] and J Elf[1]*

1. Department of Cell and Molecular Biology, Science for Life Laboratory, Uppsala University, SE-751 24 Uppsala, Sweden. 2. Division of Scientific Computing, Department of Information Technology, Uppsala University, SE–75105 Uppsala, Sweden. 3. Department of Bioengineering, University of California, San Diego, La Jolla, California 92093, USA.

*Corresponding Author

¤ Equal contribution

We have developed a method combining microfluidics, time-lapsed single-molecule microscopy and automated image analysis allowing for the observation of an excess of 3000 complete cell cycles of exponentially growing *Escherichia coli* cells per experiment. The method makes it possible to analyze the rate of gene expression at the level of single proteins over the bacterial cell cycle. We also demonstrate that it is possible to count the number of non-specifically DNA binding LacI-Venus molecules using short excitation light pulses. The transcription factors are localized on the nucleoids in the cell and appear to be uniformly distributed on chromosomal DNA. An increase of the expression of LacI is observed at the beginning of the cell cycle, possibly because some gene copies are de-repressed as a result of partitioning inequalities at cell division. Finally, observe a size-growth rate uncertainty relation where cells living in rich media vary more in the length at birth than in generation time and the opposite is true for cells living in poorer media.

## Introduction

Using time-lapsed phase-contrast and fluorescence microscopy, it is possible to monitor live bacterial cells and simultaneously quantify the expression of their highly expressed genes as the activity of introduced fluorescence reporters [1]. However, for many of its native protein species, a bacterial cell expresses only a few copies per generation [2]. In order to study processes involving these proteins, fluorescence microscopy methods sufficiently sensitive to resolve individual molecules have been developed. For instance, Yu *et al.* reported in 2006 of the use of a fast maturing yellow fluorescent protein variant, Venus [3], fused to a membrane tag, Tsr, to profile the absolute expression of the *lacZ* gene, in live *Escherichia coli* cells in its repressed state [4]. The Tsr domain immobilizes the fluorophore at the membrane so that it appears stationary for periods of 50-100 milliseconds and can be detected as a diffraction-limited spot. However, tethering to the membrane will disable molecules that rely on intracellular mobility for their function. For this reason method for counting expression events for cytoplasmic proteins have been limiting. A possible solution is suggested by the single molecule tracking experiments performed in the Xie lab, in which stroboscopic illumination pulses were used to image the transcription factor LacI-Venus non-specifically bound to DNA in live *E. coli* cells [5]. This suggests that short excitation pulses could be used also to profile the synthesis of cytoplasmic low copy number transcription factors or other proteins binding to relatively immobile intracellular targets.

Single-protein counting experiments *in vivo* reveal that isogenic cells under seemingly identical experimental conditions display considerable diversity in expression [6]. In order to confidently draw conclusions on the nature of this diversity it is necessary to sample a sufficient number of cells. Several microfluidic devices have been reported to substantially increase experimental throughput by harnessing the reproduction of bacterial cells to continuously regenerate the sample and also allowing imaging of many replicate colonies in parallel [7,8]. However, the sheer size of image data sets which can be generated in this fashion overwhelms manual analysis efforts and consequently several initiatives of automation have been undertaken [9,10]. In this study we report of a method combining microfluidics, single-molecule fluorescence microscopy and automated image analysis, enabling the study of the expression and super-resolution localization of low copy number transcription factors throughout thousands of bacterial life spans per experiment. To illustrate the performance of the method we quantify the dynamics of synthesis and intracellular localization of the lactose repressor by monitoring LacI-Venus expressed from its native promoter in live *E. coli* cells. We compare these observation with those obtained under identical conditions for cells expressing the reporter construct Tsr-Venus from the lactose permease gene, *lacY,* of the lactose operon.

## Materials & Methods

### Design, Fabrication and Use of the Microfluidic Device

The chip design was inspired by Mather *et al.* [11]. The features of the microfluidic chip used in this study were designed in three layers using AutoCAD. The layers correspond to structures of different step heights of the mold and ultimately to the different depths of the structures of the finished microfluidic device (described under "Mold Fabrication" and "Chip Fabrication"). The device contains four structural motifs; ports, channels, a chamber and traps (Fig 1a). The chamber houses three evenly spaced rows, each containing 17 traps (Fig 1a). Each trap is 40 µm x 40 µm x 0.9 µm (Fig 1b), which is bounded by two opposite walls and two open sides connecting the trap to the 10 µm deep surrounding. This geometry restricts the cells to form a monolayer colony in the focal plane while imaging. Cells close to the openings are released as the colony expands (Fig 1b). The microfluidic device is connected to media reservoirs and imaged using an inverted microscope (Fig 1c).

The master mold was fabricated using standard UV-soft lithography techniques. Three masks for microfabrication were printed in chrome. Custom formulations of SU8 Photoresist (MicroChem) were deposited on clean polished silicon wafers (University Wafer) using a spin coater. The wafers were then aligned to the mask and exposed using a mask aligner (Süss MA6). This process was repeated to deposit layers of step heights 0.9, 2.7 and 10 µm per wafer. The first layer corresponds to the trap depth of the microfluidic device; the intermediate layer enables the alignment of the first and third layer, which corresponds to the channels and ports. Each layer of the molds was measured using a stylus profilometer and inspected under a microscope before applying the next.

A master cast of the mold was made from polydimethylsiloxane, *PDMS* (Sylgaard 184, Dow Corning), using the master mold. Bubbles were removed by vacuum desiccation. The cast was cured at 80ºC for 30 minutes. One master cast contained 12 identical chip structures, which could be excised and used individually.

When fabricating each device, port holes (0.5 mm diameter) were punched out of the device cast. Debris was removed from the cast by vortexing in Ethanol. The chip cast was bonded to a coverslip (40 mm diameter, 200 µm thick, thermo-Scientific) after Oxygen/UV-plasma treatment (UVO-cleaner 42-220, Jellight Co.) for 5 minutes at 0.5 bar Oxygen pressure. The bond was stabilized by incubating at 80ºC for 10 minutes. Just prior to loading and running the device, the ports were treated with a high frequency generator (model BA 20 D, Electro-Technic Products Inc.) and the device was flooded with de-ionized water.

Gravity flow was used to control the direction and the magnitude of the flow inside the microfluidic device. The pressure gradients between the different ports of the device were established by differences in elevation relative the sample of the connected reservoirs. During loading, the seeding culture was introduced into the device through the running waste port. The cells were caught in the traps by introducing pressure waves into the tubing. Once all traps were sufficiently occupied (10-100 cells per trap), the direction of the flow in the chamber was reversed, exchanging the seeding culture with fresh medium (Fig 1a). The cells were allowed to acclimatize and grow until the traps were fully occupied (~4 hours) before imaging. The temperature of the sample was maintained at 37ºC using a custom-fitted incubator hood (OKO-lab).

**Strains and Medium**

**Two bacterial strains**, SX701 and JE116, based on *E. coli* strain BW25993 [12], were used in this study. In strain SX701, the lactose permease gene, *lacY*, was replaced with the *tsr-venus* construct [13]. Strain JE116 is based on strain JE12 [5], in which the *lacI* gene was modified to encode a C-terminal fusion of LacI and Venus. The auxiliary lactose operator site, *O3*, was replaced with the main operator sequence, *O1*, to increase auto-repression by LacI threefold. Further, in strain JE116 the downstream sequence, including the native O1, O2 binding sites as well as parts of the *lacZ* gene was removed, leaving only one specific binding site sequence for LacI-Venus molecules per chromosome copy [14].

**Cells were grown** in M9 minimal medium, with 0.4% Glucose, either with or without supplemented amino acids (RPMI1640 (R7131), Sigma-Aldrich). An overnight culture was diluted 200 times in 40 ml fresh medium and incubated for 3-5 hours (6-8 hours for cells grown without amino acids) at 37ºC and shaking at 225 rpm. During this incubation the microfluidic device was prepared. Cells were harvested into a seeding culture by centrifugation at 5000xrcf for 2.5 minutes and the pellet resuspended in 50-100 µl fresh medium. In order to prevent the cells from sticking to the surfaces of the microfluidic device a surfactant, Pluronic F108 (prod. Number 542342, Sigma-Aldrich), was added to all medium to a final concentration of 0.85g per litre.

**Microscopy & Imaging**

**Imaging** was performed using an inverted microscope (TI Eclipse, Nikon) fitted with a high numerical aperture oil objective (APO TIRF 100 x / N.A 1.49, Nikon) and external phase contrast to minimize loss of fluorescence signal. The phase contrast channel and the fluorescence channels were imaged using separate cameras, a model CFW-1312M (Scion Corporation) and a Ixon EM plus (Andor Technologies) respectively. Focus was maintained by the *Perfect-Focusing-System* of the microscope. The light source for fluorescence excitation was an Argon ion laser (Innova 300, Coherent Inc.) dialed to 514 nm for excitation of YFP-reporters in the sample. For fluorescence imaging, a slower shutter (LS6Z2, Uniblitz) was used for strain SX701 (Tsr-Venus) and a fast shutter (LS2Z2, Uniblitz) was used for strain JE116 (lacI-Venus). The fast shutter was controlled using a signal generator (AFG3021B, Tektronix) which was triggered by the Ixon camera, exposing the sample for 1 millisecond. A 2x magnification lens was used in the fluorescence emission path to distribute the point spread function ideally on the 16 µm pixels of the Ixon camera. Image acquisition was performed using *RITAcquire*, an in house GUI-based plugin for micromanager (v1.3.4.7, www.micro-manager.org). In each experiment three positions (traps) were subjected to the following acquisition program in parallel: every thirty seconds (every frame), a phase contrast image (125 ms exposure) was taken for all positions. Every three minutes (1/6 frames) for all positions, in addition to the phase contrast image, two fluorescence images (50 milliseconds exposure for SX701 and 1 ms exposure for JE116) were taken in rapid succession, followed by a bright field image (100 ms exposure) of the fluorescence channel, *i.e.* using the white-light lamp of the microscope as illumination source. This programming cycle was repeated for 1001 frames (8.3 hours). Fluorescence images were acquired in tandem to account for the effects of bleaching on molecular counting (see under "Maximum Likelihood Estimate of Synthesis"). The bright field images were acquired to allow alignment of phase contrast and fluorescence images for each frame. Our automatic method for cropping the phase images and aligning them to the fluorescent images is described in the supplementary methods.

**Cell Segmentation and Tracking**

For segmenting and tracking individual cells in the microfluidic device, we have modified and further developed existing MATLAB software, MicrobeTracker [10]. MicrobeTracker uses the position of cells in the previous frame as an initial guess and applies an active contour model [15] to fit each cell with a sub-pixel resolution boundary. In order to accurately track mobile cells over several generations, three additional supervised algorithms [16] were implemented in MicrobeTracker (supplementary methods): a *cell pole tracker* and two separate *error detectors*. The cell pole tracker is used to help the active contour model find the cell poles correctly for moving cells; otherwise this will lead to error propagation in the subsequent frames. The first error detector identifies errors made by the cell pole tracker. This is usually the result of an occasional large displacement of the cell between frames. This activates the cell tracker, which attempts to correct the segmentation of the erroneous cell. The accuracy of the cell tracker is in turn monitored by a second error detector. Any cells histories triggering this detector are terminated. In addition, a novel *division function* was added to MicrobeTracker in order to more accurately detect cell divisions for densely growing *E. coli*. Each supervised algorithm was constructed by first identifying features that efficiently discriminate between two classes, for instance, true or false cell division. In the second step, training data was extracted manually from the image sets for creating training examples for the algorithm in order to achieve accurate classification. A linear classifier [16] was used in all supervised algorithms. The algorithms, cell tracker and the classification method are described in detail in the supplementary methods. To increase the computational speed, parts of MicrobeTracker were rewritten to allow parallel computing using MATLAB's Parallel Computing Toolbox (PCT).

**Single Molecule Detection, Localization**

Fluorescent particles in the sample were detected as diffraction limited spots in the fluorescence micrographs according to the method described in [17], in which the normalized cross-correlation between the fluorescence image and an idealized optical point-spread-function (a symmetric bi-variate Gaussian function) is calculated. The standard deviation for this function is obtained experimentally by imaging and the signatures of immobilized highly fluorescent beads (data not shown). The image resulting from the correlation is transformed using the Fisher transform. A Fisher transformed Gaussian function with standard deviation corresponding to the point spread function is fitted to the fisher transformed correlation image using Levenberg-Marquardt method [18] implemented in MATLAB and the obtained parameters are used to localize each molecule with super-resolution accuracy and estimate the localization error.

**Maximum Likelihood Estimate of Synthesis**

For gene expression studies, we want to estimate how many molecules that have been newly synthesised between two fluorescence images given that there is a chance that some of the fluorophores present in the previous frame has not been bleached. We formulate this as maximum likelihood problem where there are M molecules observed in frame *i*-1 and N molecules observed in frame *i*. The number of molecules surviving bleaching, m, can be calculated by maximizing the probability

$$p(m|M,N,p,\lambda) = Bin(m,M,1-p) \cdot Po(N-m,\lambda)$$

where *Bin* is the binomial distribution and *Po* is the Poisson distribution. The maximum likelihood estimate of the number of new synthesised molecules is $n_{max}$=N-$m_{max}$ where $m_{max}$ maximizes $p(m|N,M,p,\lambda)$. The parameter *p* is the bleaching probability per fluorophore per frame and is assumed to be constant. $\lambda$ is the number of molecules synthesised between two frames.

In the special case of cell division between frame *i*-1 and *i*, where N1 molecules are found in one daughter cell and N2 in the other, the most likely number of newly synthesized molecules $n_{max}$ are calculated for both cells based on N=N1+N2. Given $n_{max}$ the most likely number of newly synthesized molecules in daughter cell 1 is the n1 that maximizes $\binom{N1}{n1}\binom{N2}{n_{max}-n1}$ since this gives the number of possible combinations of picking n1 molecules from N1 and n-n1 from N2.

**Availability**

All programs and scripts developed for this study will be made available at request.

## Results

**Throughput**

Currently, one experiment returns approximately 3000 complete cell histories from three traps imaged in parallel. The total time of expenditure is 36 hours. The manual effort of a single operator amounts to 3 hours, of which roughly 80 percent is spent prior to image acquisition. The manual work effort to acquire and analyze the images constitutes less than 2% of the total time required to complete these processes (Fig 2a). Several overlapping experiments can be performed to utilize the alternating availability of the microscope and the computational framework to further improve throughput. The number of cell histories acquired from an image series is determined in the segmentation process.

The number of cells that the program keeps track of decays over time as cells sometimes displace farther between subsequent frames than the segmentation algorithm can track them. The rate of decay varies considerably between image series, even when acquired under seemingly identical conditions (Fig 2b). Only the set of cell histories that completely cover the time from division-to-division enter the analysis (Fig 2c).

**Morphology and Growth in Microfluidic Device**

The generation time defines the growth rate of exponentially growing cells and is often used as an indicator of the health or fitness. We compare cells grown with and without amino acids in the medium (Fig 3a, red and blue respectively) and observe average generation times of 26.4 ±7.2 minutes and 46.8±17.0 minutes. Further, we observe an exponential growth of the cell length over the cell cycle (Fig 3b). In contrast to previous reports [11], we observe no obvious dependencies of the growth rate on the position the cell occupied in the trap (Fig 3c). This uniformity also holds for morphology and bacterial age, *i.e.* the number of divisions during which the oldest pole of a cell has been observed. We find that the generation times of mother and daughter cells are weakly correlated ($r$=0.27±0.02 with amino acids, $r$=0.07±0.05 without amino acids) (Fig 3d). The relation between the length at birth and the generation time of a cell history displays a correlation (Fig 3e), indicating that comparatively longer newborns complete their cell cycle faster. Although this holds qualitative for cells grown both with and without supplemented amino acids (red and blue), it is less pronounced for cells grown without amino acids. Also, cells grown with amino acids vary more in length at birth than in generation time and the opposite is observed for cells grown without supplemented amino acids. The correlation for cells with amino acids is $r$=-0.43±0.02 and without amino acids $r$=-0.28±0.04. No significant differences in growth or morphology between strains SX701 and JE116 are observed.

**Localization of Transcription Factors during the Cell Cycle**

In figure 4 we compare the intracellular localization of the reporter constructs, a) Tsr-Venus and b) LacI-Venus, over the cell cycle. A localization distribution function (Fig 4 Left column) is constructed by mapping the detected molecules to their position along the major axis of the cell (x-axis) at the time in the cell cycle they were detected (y-axis) and smoothed using a Gaussian filter. To increase synchronicity, only observations occurring in cells with generation times between 25-32 minutes and terminal lengths of 4-7 µm are included (780 for SX701 and 1176 for JE116). In figure 4a (right) we visualize the detected molecules of each construct as bi-variate symmetric Gaussian functions to create a PALM style super resolution plot of the intracellular distribution. We do not observe the typical polar localization that may be expected for Tsr (Fig 4a). This is most likely because the protein is inserted at random positions in the membrane and bleaches before reaching the Tsr clusters in the polar regions [3]. For LacI-Venus molecules (Fig 4b), we observe a tendency to cluster at positions corresponding to the nucleoids of chromosomal DNA. The number of nucleoids doubles from two, early in the cell history, to four in the later stages which is consistent with expectations for our growth conditions.

**Synthesis Dynamics of an auto-repressed Transcription Factor throughout the cell Cycle**

Figure 5 shows lineage trees of cell histories stemming from a single ancestral root of strain SX701 (Fig 5a) and JE116 (Fig 5b) with bars corresponding to the number of Tsr-Venus and LacI-Venus molecules at the times they were synthesized. The trees are pruned as cells are lost from the segmentation and or from the trap. For Tsr-Venus expressed from the *lacY* gene we observe 1.5±0.1 molecules per expression event and 1.7±0.1 events per cell cycle. For LacI-Venus 2.2±0.05 molecules per expression event and 2.5±0.04 events per cell cycle are observed. Figures 5c and 5d show the average expression rates of Tsr-Venus and of LacI-Venus molecules over the cell cycle, respectively. Both show relatively large statistical errors, especially Tsr-Venus. The cell histories with generation time 25-32 minutes and terminal length 4-7 µm are used. For strain JE116, 1418 complete cell histories and 7910 LacI-Venus molecules are observed. For strain SX701, 780 cell histories are retained from the experiment and 1176 Tsr-Venus molecules. The combination of fewer cell histories and lower expression levels leads to larger statistical uncertainty in determining the expression rate of Tsr-Venus from the *lacY* gene. However, our results indicate a greater expression rate of LacI-Venus at the beginning of the cell cycle.

**Discussion**

In this study we report of a method combining microfluidics, time-lapsed single-molecule microscopy and automated image analysis capable of monitoring the growth and absolute number of gene expression events throughout ~3000 complete individual *E. coli* life spans per experiment. Further, we demonstrate that it is possible to use a functional transcription factor, LacI-Venus, non-specifically interacting with DNA to retrieve information on both expression dynamics and super-resolution localization dynamics throughout the cell cycle. We show that the microfluidic chip provides a beneficial and stable environment for exponentially growing *E. coli* cells and a high degree of control and reproducibility. We observe a significant variability in growth rates as indicated by the generation times of individual cells. However, we find that growth rate is relatively memory-less from generation to generation. More interestingly, cells living in richer media vary more in length at birth than in generation time and that the opposite is true for cells living in poorer media. The underlying causes for this *Size-Growth Rate Uncertainty Relation* and for which range of conditions it holds are presently unclear. LacI-Venus molecules localize onto the nucleoids in the cell. It appears that non-specifically interacting transcription factors are uniformly distributed over the DNA. As expected, we find that LacI-Venus is more highly expressed than Tsr-Venus from the *lacY* gene. Our result for the latter is consistent with the findings of Yu *et al.* [4] in the number of gene expression events from the *lacZ* gene during the cell cycle. However, we observe fewer Tsr-Venus molecules per expression event (1.7±0.1 instead of 4.2±0.5). Given that *lacZ* and *lacY* are transcribed to a polycistronic mRNA, we conclude that the translation rate at the *lacY* position is two to three fold lower than that of the *lacZ* position. An increase in the expression rate of LacI-Venus is observed at the beginning of the cell cycle. We propose that this may be due to partition inequalities at cell division, in which disfavored cells replenish their transcription factor pools. The experiments confirm the highly variable nature of *in vivo* single molecule observations (Fig 5). We estimate that to obtain a 5% accuracy of the mean expression rate per minute for all points in the cell cycle would require to a total of 4000 and 16000 complete cell histories of JE116 (lacI-Venus) and SX701 (Δ*lacY::Tsr-Venus*) respectively. Sufficient observations could therefore be obtained with three additional experiments for JE116 and fifteen additional experiments for SX701. The Mather design can potentially sustain a population of bacterial cells in a state of exponential growth indefinitely. Many biological phenomena, such as the development of antibiotic resistance, occur in a small subpopulation of all cells and on longer time scales than the current longevity of an experiment using our method. Further increasing the throughput and the longevity of the method to enable the study of such phenomena represents a formidable image analysis challenge. However, to our advantage is that the performance of supervised algorithms improves and can be made more advanced as more training data accumulates. We are confident that more advanced algorithms can be implemented to increase both accuracy and speed which would make it possible to acquire an arbitrary number of cell histories from a single experiment.

## Acknowledgements

The authors would like to thank Jeff Hasty and the members of his lab for guidance in establishing the microfluidic protocols. We would specifically like to thank William Mather and Octavio Mondragon-Palomino and colleagues for designing and characterizing the functions of the trap motif used in this study. We also want to thank Christine Jacobs-Wagner and Oleksii Sliusarenko for providing the original version of MicrobeTracker. The SX701 strain was kindly donated from the lab of Sunny Xie. The JE116 strain was cloned by Prune Leroy and characterized by Petter Hammar. Further, we'd like to thank Fredrik Persson for helpful discussions on photonics and dot-detection. We are also grateful for the efforts of Intakhar Ahmad for validation work on the analysis protocols. This study was funded by grants from the European Research Council, the Swedish Research Council, Knut and Alice Wallenbergs stiftelse and Görans Gustafssons stiftelse. A.M. was funded by the Center for Interdisciplinary Mathematics (CIM), Uppsala University

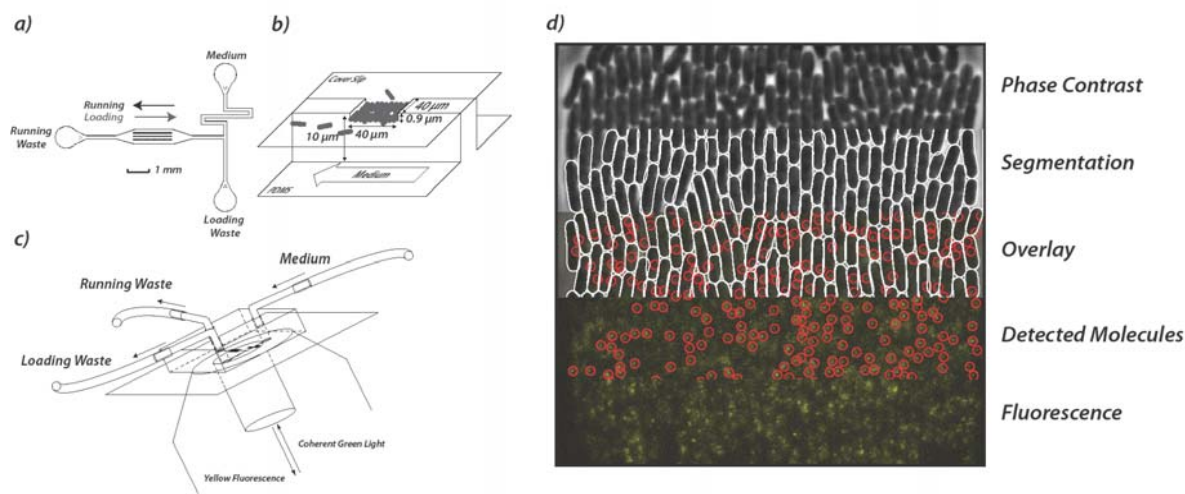

**Figure 1.** The Experimental Setup and Data Processing. a) The microfluidic device has three ports designated for medium, running waste and loading waste. The chamber houses three rows, each containing 17 traps. The direction of the flow through the chamber is alternated between the loading and running phase of the experiment. The cells are introduced from the running waste and are caught in the traps. b) Each trap is a 40 µm x 40 µm x 0.9 µm compartment which is bounded by two rigid walls and two openings. Cells which reach the openings are released from the traps into the 10 µm deep surrounding. c) The device is connected to reservoirs at the ports and imaged using an inverted microscope. The various parts of the microfluidic chip are not drawn in scale. d) Data Processing. Cells are detected and segmented from the phase contrast image (top). Molecules are detected within the fluorescence images (bottom). The coordinates from the detected molecules and cells are used to map molecules to cells (middle).

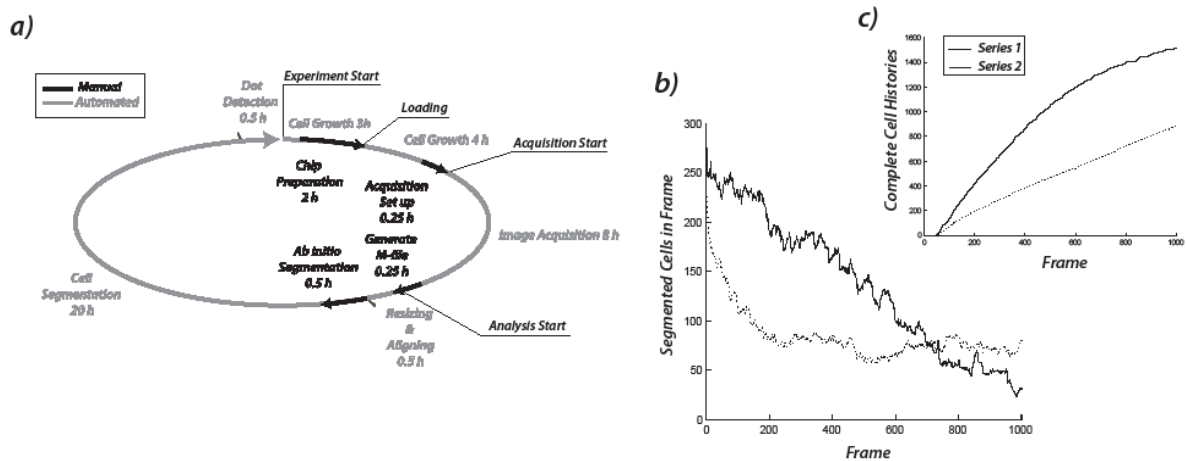

**Figure 2.** Throughput of the method. a) Time distribution for an experiment. The minimum time expenditure for an experiment is around 36 hours. Currently the method generates around 3000 complete cell generations per experiment. The protocol contains mostly automated, *i.e.* steps which require no attention from the operator. The manual time expenditure accounts for less than 10 percent of the total time and less than 2 percent of the analysis time. b) The number of segmented cells remaining at different frames during the analysis of two different time series. Due to large movements in the cell colony, the loss of correctly segmented cells varies between series. The nature of the decay is observed to depend on the pattern in which the cells grow in the trap, for which no sufficiently accurate prediction model has yet been found. c) The integrated numbers of cells acquired from cell division to cell division, *i.e.* the number of complete cell histories, for the two series in *b*.

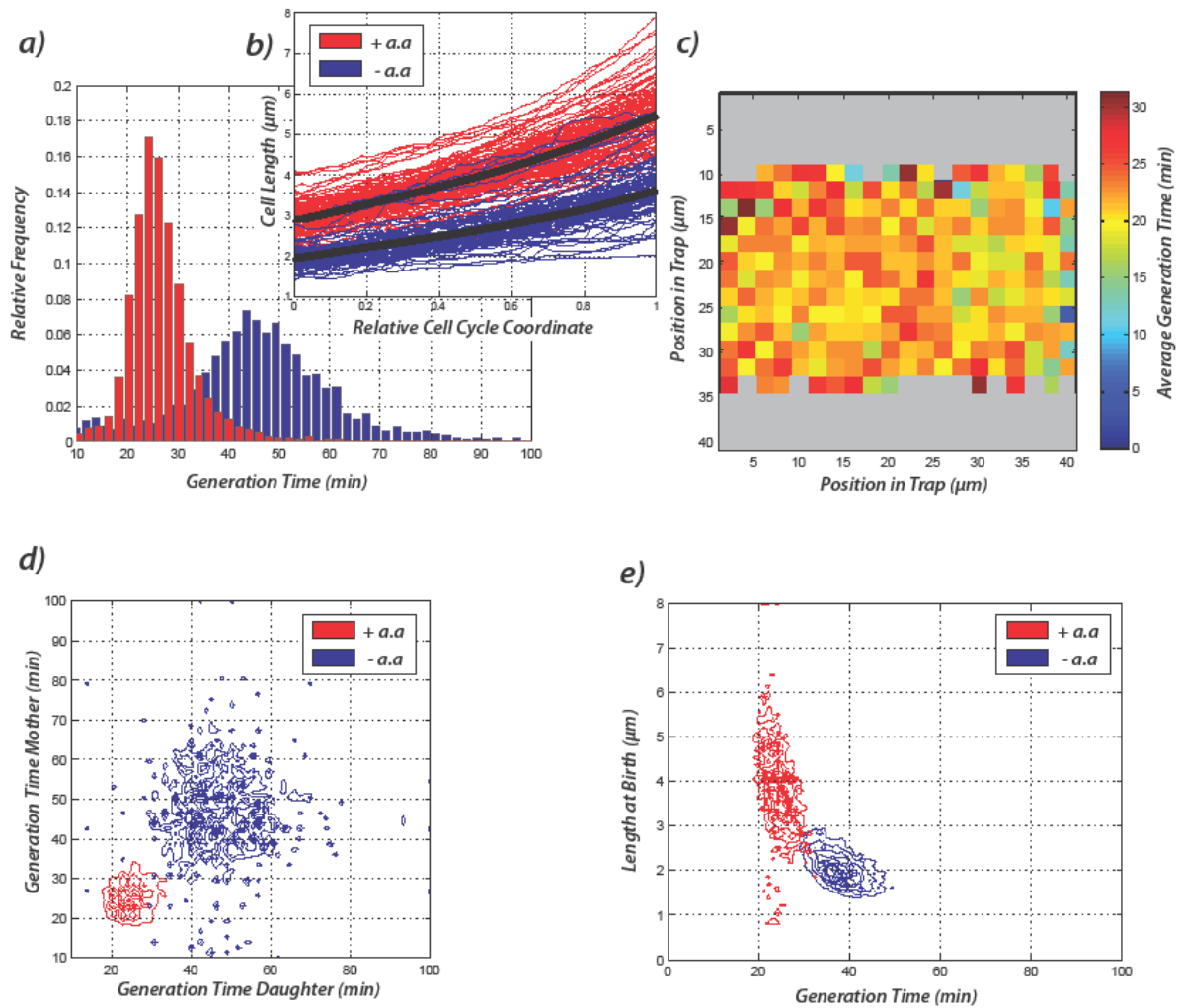

**Figure 3.** Morphology and Growth of Cells in the Microfluidic Device. a) The distribution of generation times of cells grown with (red, n=6755) and without (blue, n=2298) supplemented amino acids. b) Cell length as a function of relative cell cycle coordinates, *i.e.* time from birth normalized by the generation time, for randomly selected cells with (red, n=6755) and without (blue, n=2298) supplemented amino acids. c) Growth rate as indicated by average generation time over the geometry of the trap for cells growing with supplemented amino acids (n=6755). The figure is oriented so that the outlets of the trap are on top and bottom (see Fig 1b). d) Joint distribution of generation times for daughter and mother cells with (red, n=6755) and without (blue, n=2298) supplemented amino acids. 3 e) Joint distribution of generation time and cell length at birth for cells with (red, n=6755) and without (blue, n=2298) supplemented amino acids.

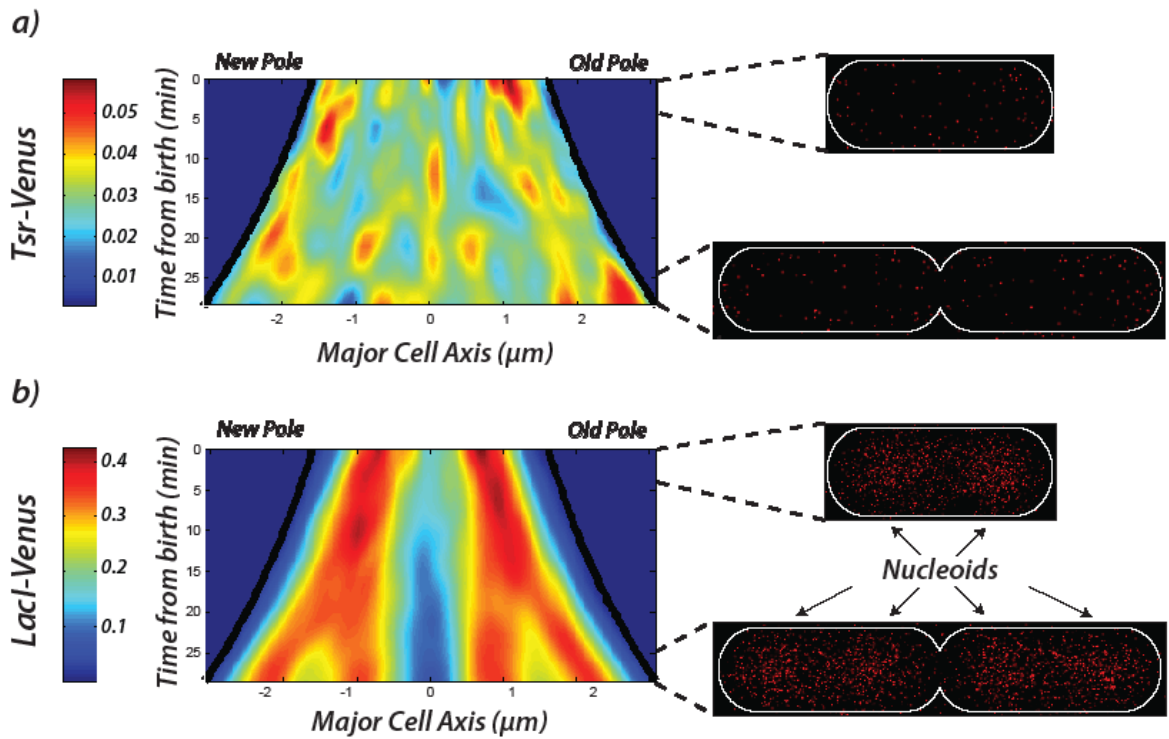

**Figure 4.** Intracellular Localization of Molecules over the Cell Cycle. In the left column are the observed molecule densities projected onto the major axis of the cell as a function of time from birth to division for a) Tsr-Venus and b) LacI-Venus. The units are the average number of molecules per minute and µm. The black lines at the edge indicates boundary of cell at the apex of the cell poles and expands as the cell grows. The right column shows the localization of individual a) Tsr-Venus and b) LacI-Venus molecules with super-resolution accuracy for early and late stages of the cell cycle.

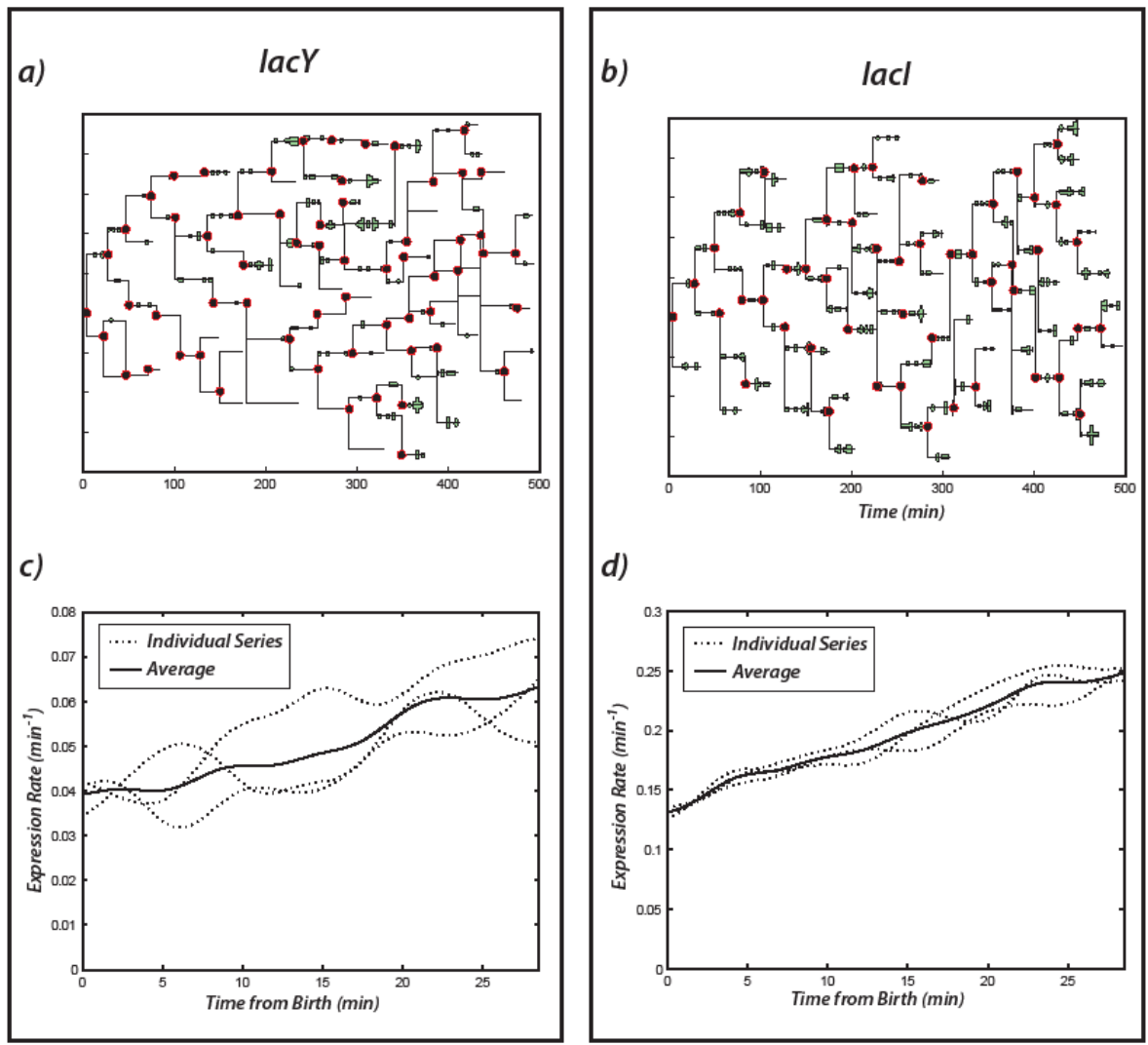

**Figure 5.** The Rate of Gene expression over the Cell Cycle. Representative lineage trees for strains SX701 (a) and JE116 (b) stemming from one ancestral root respectively. The absolute number of newly synthesized molecules expressed from the *lacY* and *lacI* gene are indicated as green bars at the time they are detected. The average expression rates from the *lacY* (c) and *lacI* (d) genes over the cell cycle. Solid lines show the average of all three series for each construct. The average of the individual series is shown as dotted lines as an indication of the uncertainty in determining the mean.

# Supplementary methods

**Cropping and Aligning Images**

In raw phase contrast images, the trap only covers a fraction of the image. In the images used for the segmentation and tracking, the images were pre-processed in order to extract the trap. This was done by correlating a binary image containing a box of a size corresponding to the chip in the raw phase image. The size of the cut-out was reduced from 860x860 pixels to 500x500 pixels in order to increase speed of our algorithms. The cut-out from the phase image was correlated with the corresponding bright field image taken in the fluorescence channel in order to get an alignment with the precision of approximately one pixel. Vertical interference patterns of light that appears in the raw phase images also needed removal. This is done by calculating a background image averaged in time as well as in the horizontal direction. The background image is subtracted from each individual phase image.

**Division Function**

The intensity landscape of a phase contrast image closely resembles a saddle node around the point where the division occurs, where a local maximum is found in the direction of the major axis and local minima in the direction of the minor axis. The original division function of MicrobeTracker classified division events solely by the magnitude of a local maxima relative its surroundings in the image. This method often mistakenly identifies unrelated intensity variations in densely growing E. coli as division events. Consequently the function was further developed to increase robustness and accuracy. The new division function applies a linear discriminant (described under "Classification") to a space spanned by five features in order to classify a potential division event as either divided or non-divided. Test positions along the cell profile are identified using the original division function [1] and around each position a reference coordinate system of 11x11 points is established. The coordinate system uses an orthonormal base were the first direction corresponds to the direction between the cell poles and the second direction corresponds to the cell width. The distance between each point was one pixel, and a typical cell width in our images was 15 pixels. A test image was calculated from this coordinate system using bi-linear 2D interpolation. The features extracted from these images are the relative magnitude of the local maxima as in the original division function, the two Eigen-values to the Hessian matrix and the scalar products of the test image with two templates of a true division event and a false division event. The templates were constructed prior to analysis by averaging corresponding 11x 11 cut-outs from manually assembled training sets consisting of 82 true divisions and 259 falsely identified division events respectively.

**Pole Searcher**

Cells continuously shift their position between the previous and the current frame. It was noticed that accurately identifying the cell poles was the most critical factor when tracking a cell and for this reason a method based on tracking the poles between frames was developed. The features of the pole searcher are the 20 principal components of the pixel intensities in a training set of 52370 manually segmented 15x15 pixel cell pole examples and an equal number of examples at random positions. A linear Bayesian probabilistic model was created from the training data (see Classification). When searching for the poles the method samples the surroundings of the previous poles stochastically according to a 4D Gaussian distribution for the spatial coordinates, cell length and an orientation angle.

This sampling and testing is done pairwise for both poles. The expectation value for the cell length is estimated using the Euler forward method and a finite difference approximation of the derivative. The expectation value of the movement of the entire cell is estimated by the optical flow method [2]. The expectation value of the orientation angle is set to the angle in the previous image. The standard deviations of these distributions are estimated from the training data. Test points are sampled from this 4D distribution and for each test point, a pair of 15x15 reference coordinate systems are established, corresponding to guesses of where the cell poles are localised in the current image. Sample images are calculated from the reference coordinate systems using 2D bi-linear interpolation. The probabilistic model described in the "Classification" section is used for calculating a probability for each pair of sample images to be correctly aligned with the cell poles. The weighted mean of all samples is used as an estimate of the new pole positions, length and angle.

**First Error Detector**

Segmentation errors will propagate throughout the series and also cause neighbouring cells to be erroneously segmented. Therefore, it was necessary to develop methods of detecting errors as they were made by the segmentation algorithm. Different indicators of errors were used to create features for a classifier to detect incorrectly segmented cells. For training, 135 correctly segmented examples and 41 incorrectly segmented examples were used. The features used were 1) the relative difference in pixel intensity inside the cell contour, 2) the overlap between cells relative to the cell area, 3) cell movement between the current and previous frame and 4) angular movement between the current and previous frame.

**Cell Tracker**

In order to increase the number of cell generations a cell tracker was developed in order to save cells lost due to the first error detector. A coordinate system of 44x17 points was created in order to describe a cell in a standardised framework, with equal number of points independent of cell length. The first and last 8x17 points of the coordinate system were used for describing the cell poles. The intermediate 28x17 points cover the remaining cell with 28 equidistant ribs along the cell profile. The cell was tracked in a given frame searching for the minimal Mahalanobis distance, *d*, defined by

$$d^2 = (x_1 - x_2)^T \Sigma^{-1} (x_1 - x_2), \tag{1}$$

where $x_1$ is the cell in the previous frame and $x_2$ in the current frame. $\Sigma$ is the covariance matrix calculated from 13880 training examples extracted in the neighbourhood of correctly segmented cells. The dimensionality of the covariance matrix was reduced by projecting the training data to the 40 first principal components. The dimension reduction was done for the purpose of removing noise in the covariance matrix due to the limited number of training examples. The search was performed by sampling from a 3D Gaussian distribution of positions and angles. The cell length was kept fixed as the same as in previous frame. The search space was reduced by not allowing a relative overlap with neighbouring cells of more than 25% of the cell area.

**Second Error Detector**

Also the cell tracker occasionally makes errors. A second error detector was therefore developed in order to prevent these errors from propagating. It uses a set of five features that are different from those used in the first error detector. These features are: 1) the relative difference in the number of pixels above the threshold given by Otsu's method [3] inside the cell contour. 2) The sum of the cell profile's absolute deviation from the straight line that goes through both cell poles. This feature was used as a measure of overall curvature. 3) The difference in the number of edge pixels inside the cell profile. The edge pixels were defined as pixels having the sum of second derivatives in the image, in x and y directions, above a pre-defined threshold. 4) Differences in cell lengths between the two frames. 5) The cross correlation of the cell images in the current and the previous frame. The cell images were interpolated from the previously mentioned 44x17 points cell coordinate system.

**Classification**

For all algorithms described above except the division function a Bayesian probabilistic model was used for classification between two classes denoted $C_1$ and $C_2$. For example, the classes may denote whether a sample image is on the pole or off the pole. The probabilistic model can be derived from Bayes theorem using the assumption that the features for both classes have a multivariate normal distribution [4]. The probability of the input data $\boldsymbol{x}$ belonging to class $C_1$ is given by

$$p(C_1 | \boldsymbol{x}) = \sigma(\boldsymbol{w}^T \boldsymbol{x} + w_0) \tag{2}$$

The vector **w**, sometimes referred to as the linear discriminant, is calculated as

$$w = \Sigma^{-1}(\pmb{\mu}_1 - \pmb{\mu}_2) \qquad (3)$$

Where Σ is the shared within class covariance matrix of class 1 and 2, **μ**₁ and **μ**₂ are the mean feature vectors of class 1 and 2 respectively. Since class 1 and 2 may have different number of training samples, the shared covariance matrix is calculated as the maximum likelihood solution

$$\Sigma = \frac{N_1}{N}\Sigma_1 + \frac{N_2}{N}\Sigma_2 \qquad (4)$$

Where $\Sigma_1$ and $\Sigma_2$ are covariance matrixes of class 1 and 2 respectively. $N_1$ and $N_2$ are the number of training samples from class 1 and 2 respectively. N is the total number of samples. The number of features must be significantly smaller than the number of training samples in order to get a linear discriminant with high statistical precision. If the number of features is larger than the number of training samples, the covariance matrix will not be invertible. The function σ, referred to as the sigmoid function, is defined as

$$\sigma(a) = \frac{1}{1+\exp(-a)} \qquad (5)$$

The quantity $w_0$ is calculated as

$$w_0 = -\frac{1}{2}\pmb{\mu}_1^T\Sigma^{-1}\pmb{\mu}_1 + \frac{1}{2}\pmb{\mu}_2^T\Sigma^{-1}\pmb{\mu}_2 + \ln\frac{p(C_1)}{p(C_2)} \qquad (6)$$

The prior probabilities are calculated using the maximum likelihood solution $p(C_1) = \frac{N_1}{N}$. For the division function, the linear discriminant was used without a probabilistic framework. In this case the scalar product $\pmb{w}^T\pmb{x}$ is calculated and compared to a threshold to check whether the cell is divided or not.

**Parameter Estimation for Maximum likelihood Estimate of Synthesis**

Two parameters, *p* and λ, were used in the Maximum likelihood method used for estimating the number of new spots. The parameter *p* was estimated by the fraction of dots that are lost by taking one fluorescence image rapidly after another. If there are *k* extra bleaching images in between the images where dots are counted p=1-(1-p₁)$^{k+1}$, where p₁ is the fractional loss of dots per bleaching frame. λ was estimated by the number of newly synthesised molecules per generation divided by the number of frames per generation where the dots are counted. The number of newly synthesised molecules per generation was estimated by counting molecules per cell in a sample where fluorescence images are taken much more rarely than the generation time. It is equal to the average number of molecules in a newly divided cell.

**Spot Quality Estimation**

The positions of the spots were estimated with the method by Ronneberger *et al* [5]. This method is briefly described in the main article. However, the spots vary in shape and quality and we only want to take into account the most significant spots. Therefore, an objective quality criterion was needed in order to reject spots with a quality below a certain threshold. This was done with linear regression according to the model

$$Y = \mathbf{1}\beta_1 + X\beta_2, \tag{7}$$

where **Y** is the 11x11 pixel cut out from the fluorescence image at the position of the spot, **1** is a matrix of the same size with each matrix element equal to 1 and **X** is a discretized 2D Gaussian function. The 2D Gaussian function has a mean estimated with the method by Ronneberger *et al* and a standard deviation corresponding to the point spread function. In general, the mean of the 2D Gaussian does not coincide with the central pixel in the 11x11 coordinate system. In the linear regression model, the coefficient $\beta_1$ corresponds to background fluorescence and $\beta_2$ to the intensity of the Gaussian spot. The Z-score was calculated for the $\beta_2$ coefficient and we test that $\beta_2>0$. In this work, spots with a Z-score below 6 were rejected.

**Reduction of Memory Footprint**

As the size of datasets that can be processed with MicrobeTracker increases, so does the memory footprint. This eventually becomes a critical issue when the managing of cell data starts to hamper performance. Originally, the cell list that contains all data of individual cells, including their contours, was stored as an array (a "cell array" in MATLAB terminology) where the number of elements for a particular frame grew exponentially with the number of frames. A new API was developed to store the cell list in a more compact way that grows linearly with the number of frames. To further reduce the memory usage by a factor of approximately 0.7 all floating-point data were stored in single precision, with no significant impact on the accuracy of the algorithms. Conversion from the old to the new format is automatically done when data is loaded, making our version of MicrobeTracker backwards compatible. The API for the new format was spliced off from the main code so that it may be used by external programs that uses the cell lists.

**Stubs and Iterative Segmentation**

To progress the transition towards fully automated segmentation the concept of *stubs* were introduced and implemented in MicrobeTracker. Whenever a cell is manipulated such that its data is invalidated on subsequent frames, e.g., by manually forcing it to split or by refining its contour, it is turned into a stub. Stubs are easily distinguished in the GUI and are exempt from further automatic segmentation unless MicrobeTracker is explicitly told to process stubs. This enables automated bookkeeping of cells with invalidated downstream data stemming from manual editing, but the stubs API was also integrated with the error detectors to exempt cells that the latter identify as erroneously segmented from further processing; i.e. cells caught in the error detector turn into stubs. The present design of the first error detector renders false positives that wrongly exempt cells from processing at subsequent frames. To compensate for that shortcoming all newly formed stubs are optionally and automatically segmented one or several times to see if a sensible cell model can be constructed. A single iteration has proven to take care of most false positives, extending the expected number of frames that a cell can be automatically segmented before MicrobeTracker loses track of it. It should be emphasized that this effect will be reduced or eliminated if the error detection is further improved such that it gives less false positives. For this work, the manual effort was only done in the initial frame.